\documentclass[prd,twocolumn,showpacs,floatfix,nofootinbib]{revtex4}
\usepackage{bm}
\usepackage{amssymb}
\usepackage{graphicx}
\usepackage{epstopdf}

\begin{document}
\title{The late-time tails in the Reissner-Nordstr\"{o}m spacetime
revisited}
\author{Carl J.~Blaksley$^{1}$ and Lior M.~Burko$^{1,2}$}
\affiliation{$^1$ Department of Physics, University of Alabama in Huntsville, Huntsville, Alabama 35899, USA\\
$^2$ Center for Space Plasma and Aeronomic Research, University of Alabama in Huntsville, Huntsville, Alabama 35899, USA}
\date{
%draft of 
October 12, 2007}
\begin{abstract}
We propose that the late-time tail problem in the Reissner-Nordstr\"{o}m (RN) spacetime is
dual to a tail problem  in the Schwarzschild spacetime with a different initial data set: at a fixed observation point the asymptotic decay rate of the fields are equal. This duality is used to find the decay rate for tails in RN. This decay rate is exactly as in Schwarzschild, including the case of the extremely-charged RN spacetime (ERN). The only case where any deviation from the Schwarzschild decay rate is found is the case of the tails along the event horizon of an ERN spacetime, where the decay rate is the same as at future null infinity. As observed at a fixed location, the decay rate in ERN is the same as in Schwarzschild. We verify these expectations with numerical simulations.
\end{abstract}
\maketitle

\section{Introduction and summary}
Late-time tails in black hole spacetimes have been studied extensively 
since Price's seminal analysis \cite{price}. Complete understanding is
available for the Schwarzschild spacetime --- where it was also found in
fully nonlinear numerical simulations \cite{gundlach,burko-ori} --- and
much progress has been made recently also in understanding the tail
problem for Kerr black holes \cite{burko-khanna,gleiser}, and
even for spacetimes without any assumption of symmetry (but with a globally weak field) \cite{poisson}. 
The tail of spherically-symmetric, electrically-charged, static black
holes --- namely Reissner-Nordstr\"{o}m (RN) black holes --- was studied first by
Bi\v{c}\'{a}k for the case of scalar perturbations (for both the non-extreme and extreme cases) in \cite{bicak} and for the case of coupled electromagnetic and gravitational (linearized) perturbation in \cite{bicak1} (see also \cite{bicak2}) and revisited by Gundlach, Price, and Pullin (GPP) \cite{gundlach1}. Price's law has been proved rigorously in the Schwarzaschild case and a self-gravitating 
scalar field by Dafermos and Rodnianski \cite{dafermos}.  

The late-time tails in black hole spacetimes are understood as a result of
the asymptotic form of the effective potential at great distances from the
black hole, and the exponential drop off of the potential close to the
event horizon. It is the scattering of the waves off the effective
potential at great distances that is responsible for the creation of the
tails. (This picture applies also for the case of a Kerr spacetime, but
this simple behavior is masked by an intricate mode coupling effect. See
\cite{burko-khanna}.) Indeed, it was implied first by Bi\v{c}\'{a}k in Ref.~\cite{bicak} and then argued explicitly 
in greater detail by GPP in Ref.~\cite{gundlach1} that because in RN the effective potential has the same form at great
distances as in Schwarzschild, the decay rate of the tails in RN must be
the same as in Schwarzschild. It was further demonstrated numerically in
Ref.~\cite{gundlach1} that this is indeed the case. (This argument is
further reinforced by the analysis in Ref.~\cite{ching}.) 

The GPP argument suggests that all spacetimes that share the same
asymptotic form of the effective potential, have the same decay rate for
the late-time tails. The possible exception of the late-time tails in the extreme
RN (ERN) spacetime is therefore intriguing. Specifically, Bi\v{c}\'{a}k
argued in Ref.~\cite{bicak} that the exponent of the tails (as observed
in a fixed observation point) in the ERN spacetime for initial data with
an initial static moment $\ell$ was $\ell+2$, whereas in Schwarzschild and
(non-extreme) RN it is $2\ell+2$. The asymptotic form of the effective
potential is the same in those three spacetimes. (They differ only to
$O(r_*^{-3})$, $r_*$ being the tortoise coordinate. To 
$O(\ln (r_*/M)\, r_*^{-3})$ the three effective potentials are
identical).  Applied naively, the GPP argument suggests that the same
exponent for the tails should be observed for all three cases. 

However,
it turns out that the effective potential in ERN is very different from
the effective potentials in Schwarzschild or (non-extreme) RN near the
black hole: instead of dropping off exponentially (in $r_*$, for
large and negative values of $r_*$) towards the event horizon, the
effective potential in ERN is effectively centrifugal asymptotically close
to the event horizon. Moreover, the deviations from centrifugality have
the same leading-order form as the deviations from centrifugality at great
distances from the event horizon (large and positive $r_*$). That is, the
effective potential is asymptotically symmetric \cite{bicak}, such that the tails at 
$r={\rm const}$ are expected to have contributions both from scatterings
at great distances, and from scatterings very close to the event
horizon. (In that sense, the event horizon is equivalent to future null
infinity. In fact, the close analogy of the event horizon to future null
infinity is even deeper: the tail along the event horizon of ERN turns out
to be the same as along future null infinity (see below).) 

Can the contributions to the tails coming from the close vicinity of the
event horizon in ERN overwhelm the contributions to the tails coming from
great distances, or interfere with them to create tails with the indices predicted in \cite{bicak}? In this paper we study this issue, and show that with
adaptations, the GPP argument is applicable also to the ERN spacetime. We
then present also numerical simulations of tails in ERN. These simulations indicate that the fall-off of these tails  in the ERN case 
is the same as in the non-extreme case, i.e., quicker than that predicted in Ref.~\cite{bicak}.

The organization of this paper is as follows: In Section \ref{sec_pot} we describe the scattering problem, and discuss the asymptotic symmetry of the effective potential. This symmetry was found in \cite{bicak}, and is presented here for completeness. In Section
\ref{secondary_waves} we argue that in both Schwarzschild and ERN the tails
along $r={\rm const}$ result from the secondary waves (first
scattering). Tertiary waves (second scattering) or higher-order waves 
that are present at late times originate only from scatterings at 
small values of $|r_*|$; scattering at great distances leads only to
secondary waves. As it is the scattering at great distances which leads to
the tails, we argue that it is the secondary waves which are responsible
to the tails. 
%Notably, Ref.~\cite{bicak} argues that primary waves (no scattering) dominate in the tails. 
In Section \ref{com} we discuss the tails in ERN when the
initial data have compact support. We argue that the tail problem along $r={\rm const}$ in ERN is dual to 
a tail problem in Schwarzschild (for a different, but related, initial value problem); as the 
latter is well understood, we may predict the power-law indices for ERN tails. The meaning 
of duality here is as follows: while the value of the field at a particular event is different in the two spacetimes (because the  intermediate effective potentials of the two dual spacetimes are not identical), the scattering dynamics from regions of spacetime that contribute to the formation of the tail is 
similar, so that the same power-law indices for the tail are expected. We emphasize that the duality 
argument pertains only to the tail at a fixed observation point (along $r={\rm const}$).  
Then, in Section \ref{ism} we consider the tails in
ERN with an initially static moment for the initial data. We again show
that this problem is dual to another tail problem in Schwarzschild, and use
this duality to find the decay rate of the tails in ERN. Finally, in
Section \ref{sec_num} we present numerical simulations of tails in ERN which
are in full agreement with our expectations based on the arguments brought
is Sections \ref{com} and \ref{ism}. We describe the numerical code and the convergence tests done in Appendix \ref{app}.

\section{The ERN effective potential}\label{sec_pot}

The ERN metric is given by
\begin{equation}
\,ds^2=-\left(1-\frac{M}{r}\right)^2\,dt^2+\left(1-\frac{M}{r}\right)^{-2}\,dr^2+r^2\,d\Omega^2\, ,
\end{equation}
where $r$ is the regular radial Schwarzschild coordinate defined so that spheres of radius $r$ have  surface area of $4\pi r^2$, and $\,d\Omega^2$ is the line-element of the unit 2--sphere. 
In terms of the Regge--Wheeler `tortoise' coordinate $r_*$, defined by 
\begin{equation}\label{rstar}
\frac{\,dr_*}{\,dr}=\left(1-\frac{M}{r}\right)^{-2}\, ,
\end{equation}
the wave scattering problem is governed by the wave equation 
$$-\frac{\,\partial^2\psi}{\,\partial t^2}+ \frac{\,\partial^2\psi}{\,\partial {r_*}^2}-V[r(r_*)]\,\psi=0\, ,$$
where the ERN effective potential, for a multipole $\ell$, is given by
\begin{equation}\label{eff_pot}
V(r)=\left(1-\frac{M}{r}\right)^2\,\left[\frac{2M}{r^3}\left(1-\frac{M}{r}\right)+\frac{\ell (\ell+1)}{r^2}\right]\, .
\end{equation}
For a scalar field $\phi$, the dimensionally--reduced field $\psi$ is defined by $\phi=r\,\psi$. For field spins other than $s=0$, the effective field $\psi$ is interpreted appropriately \cite{bicak1,bicak2}.

To find the {\em asymptotic} effective potential, integrate Eq.~(\ref{rstar}) to find 
$$r_*(r)=r+2M\ln\left(\frac{r}{M}-1\right)-\frac{M^2}{r-M}+{\rm const}\, .$$
Denoting
\begin{eqnarray}
\rho &:=& \frac{r-M}{M}\\
R &:=& \frac{r_*-M}{M}\, ,
\end{eqnarray}
and choosing the integration constant, 
the relation between the radial coordinates is
\begin{equation}\label{R_rho}
R=\rho+2\ln\rho-\frac{1}{\rho}\, .
\end{equation}

In order to find $V(r_*)$ as $r_*\to\pm\infty$, consider two cases: Case a) $r\gg M$ ($\rho ,R\gg 1$), for which 
$R\sim \rho +2\ln\rho$ (neglecting the term $-1/\rho$ in Eq.~(\ref{R_rho})), and Case b) $0<r-M\ll M$ ($0<\rho\ll 1$, $R\ll -1$, $|R|\gg 1$), for which 
$R\sim 2\ln\rho-\rho^{-1}$ (neglecting the term $\rho$ in Eq.~(\ref{R_rho})).

\subsection{Case a: $r\gg M$}
In this case 
$$R\sim\rho+2\ln\rho, ,$$
which can be solved as
\begin{equation}
\rho\sim 2\, W\left(\frac{1}{2}e^{R/2}\right)\, ,
\end{equation}
where $W(x)$ is the Lambert $W$--function\footnote{For the Schwarzschild spacetime in the usual Schwarzschild coordinates, the `tortoise' coordinate is defined by $\,dr_*=\,dr/(1-2M/r)$. Defining ${\tilde\rho} := r/(2M)-1$ and ${\tilde R} := r_*/(2M)-1$, and choosing the integration constant, the two radial coordinates are related by $\tilde R=\tilde\rho+\ln\tilde\rho$. In terms of the Lambert $W$--function we may write $\tilde\rho=W[\exp(\tilde R)]$, or
$$r(r_*)=2M+2M\,W\left(e^{\frac{r_*-2M}{2M}}\right).$$} \cite{W}, defined by the inverse function of $W(x)\, e^{W(x)}=x$ (i.e., the (real) function $W(x)$ which solves this equation). As by definition $\rho>0$, the principal branch of the $W$--function (denoted $W_0(x)$) is chosen, and one does not have branch ambiguity.  
 For $x\gg 1$, $W(x)\sim\ln x-\ln\,\ln x+\cdots$, so that 
\begin{eqnarray}
\rho &\sim& 2 \ln \left( \frac{1}{2}\, e^{R/2}\right)-2 \ln\left[\ln\left(\frac{1}{2}\, e^{R/2}\right)\right]+\cdots \nonumber \\
&\sim & R-2\ln\frac{R}{2}+\cdots  \, ,
\end{eqnarray}
or 
\begin{equation}
r(r_*)=r_*-2M\ln\frac{r_*}{2M}+\cdots\, .
\end{equation}
Substituting in Eq.~(\ref{eff_pot}), one finds ($r_*\gg M$):
\begin{equation}
V(r(r_*))=\frac{\ell (\ell+ 1)}{r_*^2}+4M\frac{\ell (\ell+1)}{r_*^3}\,\ln\frac{r_*}{2M}+O(r_*^{-3})\, .
\end{equation}

\subsection{Case b: $0<r-M\ll M$}
In this case 
$$R\sim -\frac{1}{\rho}+2\ln\rho \, ,$$
which can be solved as
$$\rho\sim e^{-|R|/2}\, \exp\left[W\left(\frac{1}{2}e^{|R|/2}\right)\right]\, .$$
Using the definition of the Lambert $W$--function,  
\begin{eqnarray*}
\rho &\sim & e^{-|R|/2} \times\frac{1}{2}\frac{e^{|R|/2}}{W\left(\frac{1}{2}\, e^{|R|/2}\right)}\\
 &\sim & \frac{1}{2\, W\left(\frac{1}{2}\, e^{|R|/2}\right)}\, ,
\end{eqnarray*}
so that
\begin{eqnarray}
\rho &\sim & \frac{1}{2\times\frac{|R|}{2}\left(  1-\frac{2}{|R|}\ln\frac{|R|}{2}+\cdots  \right)}\nonumber \\
&\sim & \frac{1}{|R|}+\frac{2}{R^2}\ln\frac{|R|}{2}+\cdots\, ,
\end{eqnarray}
and
\begin{equation}
r(r_*)=M-\frac{M^2}{r_*-M}\left[  1-\frac{2M}{r_*-M}\ln\left|\frac{r_*-M}{M}\right| +\cdots \right]\, .
\end{equation}
Substituting in Eq.~(\ref{eff_pot}), one finds ($r_*\ll -M$):
\begin{equation}
V(r(r_*))=\frac{\ell (\ell+ 1)}{r_*^2}-4M\frac{\ell (\ell+1)}{r_*^3}\,\ln\frac{|r_*|}{2M}+O(r_*^{-3})\, .
\end{equation}

\begin{figure}[htbp]
   \includegraphics[width=3.4in]{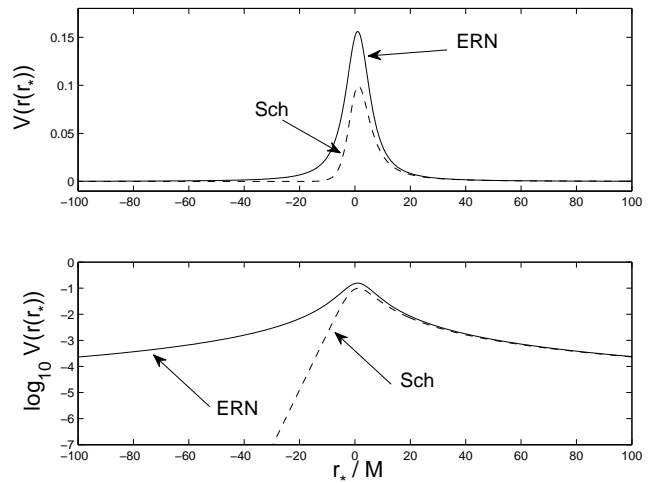} 
 \caption{The effective potential for $\ell=1$ in ERN (solid curve) and Schwarzschild (dashed curve), as a function of $r_*$.}
\label{pot}
\end{figure}

Combining the results for Cases a) and b), one finds ($|r_*|\gg M$)
\begin{equation}
V(r(r_*))=\frac{\ell (\ell+ 1)}{r_*^2}+4M\frac{\ell (\ell+1)}{|r_*|^3}\,\ln\frac{|r_*|}{2M}+O(r_*^{-3})\, .
\end{equation}
The last expressions shows explicitly that the effective potential is asymptotically symmetrical under $r_*\leftrightarrow -r_*$. Notably, the asymptotic effective potential for ERN coincides with the Schwarzschild asymptotic effective potential (for $r_*\gg M$). The same effective potential is also given in \cite{bicak}. We add that in terms of the Schwarzschild radial coordinate $r$, for $r\gg M$, 
$$V(r)=\frac{\ell (\ell +1)}{r^2}-2M\frac{\ell^2+\ell -1}{r^3}+O(r^{-4})$$
is the same in Schwarzschild as in ERN. The ERN and Schwarzschild effective potentials, for $\ell=1$, are shown in Fig.~\ref{pot}. 

The key step in finding the asymptotic effective potential was finding $r(r_*)$. The latter may be found without invoking the Lambert $W$--function by an iterative solution, in which one substitutes $r_*(r)$ into
the equation $r=r_*+g(r)$. After each iteration, the right hand side will be a combination of terms in $r_*$ and in $r$,  but the latter terms become smaller with each iteration, and may be neglected.

\section{The late time tails result from secondary waves}
\label{secondary_waves}

\begin{figure}[htbp]
  % \centering
   \includegraphics[width=3.4in]{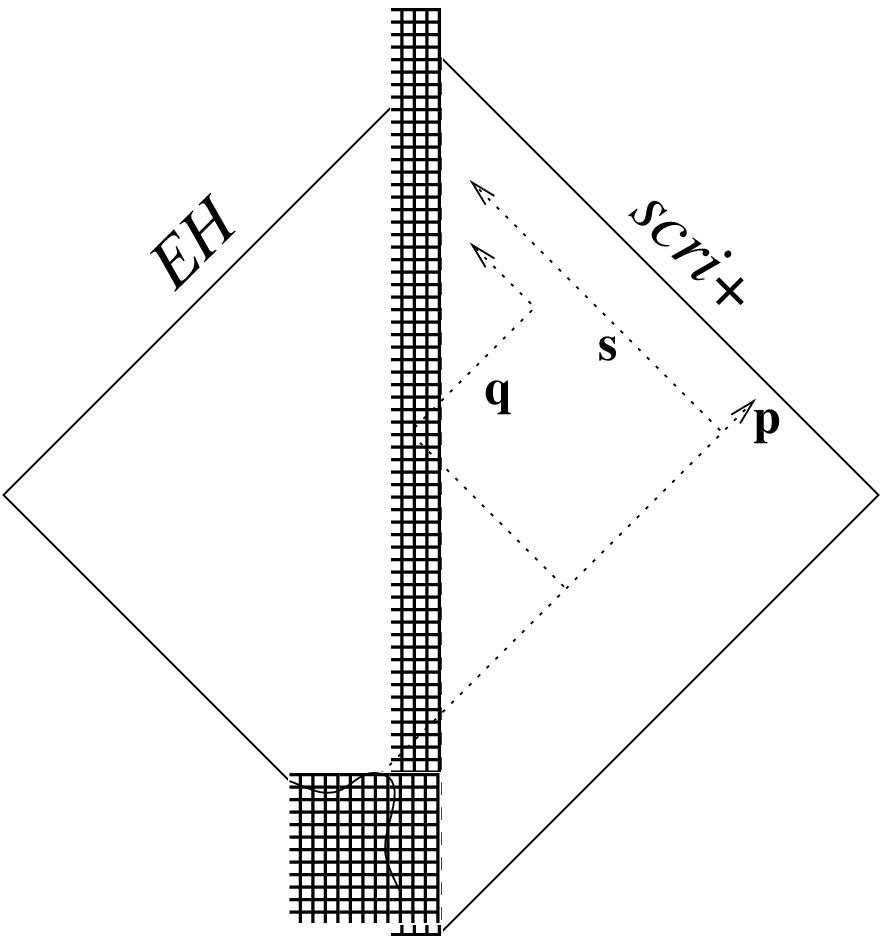} 
 \caption{A schematic diagram of scattered waves. The initial pulse in this
Figure is initially outgoing (and is therefore shown on the incoming leg of the 
characteristic hypersurface), with the primary waves (p) arriving to
scri$_+$ in the Figure. The waves which cross the worldline of an observer
at $r={\rm const}$ at late times are scattered waves: the secondary waves
(s) undergo one scattering event, and quaternary waves (q) undergo three
scattering events. This Figure only displayes the waves approaching the
observer from the right, hence no tertiary waves are shown. The abbreviation ``EH" 
stands for the event horizon.}
\label{scattering}
\end{figure}

Consider an initial perturbation field of compact support. (We assume
compact support here without loss of generality in order to separate at
late times between primary waves and scattered waves, as we are interested
here only in the latter.) The part of the radiative field at later times
which propagates along the geometrical optics rays is the primary
waves. These waves scatter off the effective potential. The waves which
scatter just once are the secondary waves, these which scatter twice are
the tertiary waves, and so on. (See Fig.~\ref{scattering}.)

%\begin{figure}
%\input epsf
%\epsfxsize=8.5cm
%\centerline{ \epsfxsize 8.0cm
%\epsfbox{scattering.eps}}

In this Section we show that in both the Schwarzschild and the ERN cases
the late-time tails are generated by the secondary waves, and higher-order
waves are irrelevant for their generation. The tails in Schwarzschild are generated by the
scattering of waves off the effective potential at great distances, i.e.,
it is the part of the effective potential at $r_*\gg M$ which is
responsible for the generation of the tails. In order to study which order
of the waves (i.e., secondary or higher) is relevant for the generation of
the tails, we write the effective potential in the form 
\begin{equation}
V_{\rm eff}(r_*)=\frac{\ell(\ell +1)}{r_*^2}+\hat{V}(r_*)\, ,
\end{equation}
where the curvature potential 
\begin{equation}
\label{hatV}
\hat{V}(r_*)=A\frac{4M\ell(\ell +1)}{r_*^3}\, \ln\frac{r_*}{2M}+O(r_*^3)
\, .
\end{equation}
Here, $A$ equals unity for the Schwarzschild effective potential. (We
parametrize here the curvature potential with $A$ for the analysis of the
order of the waves below.) 

Secondary waves are those which scatter just once. Therefore, secondary
waves are expected to be proportional to $A$. Therefore, if we study the
tails as we vary the value of $A$ in Eq.~(\ref{hatV}), we can determine
whether the tails are secondary waves (linear dependence on $A$), or
include higher-order contributions (deviations from the linear dependence
on $A$). Primary waves are independent of $A$. In practice, we write the {\em toy} potential as
\begin{eqnarray}
V(r_*)=\left\{\begin{array}[c]{l}
\left(1-\frac{2M}{r}\right)\left[\frac{\ell(\ell +1)}{r^2}+
\frac{2M}{r^3}\right] \;\;\;\;\; r<r_0
\\ 
\frac{\ell(\ell +1)}{r_*^2}+
A\frac{4M\ell(\ell +1)}{r_*^3}\, \ln\frac{r_*}{2M}  \;\;\;\;\; r>r_0
\end{array}\right. \, ,
\end{eqnarray}
that is, at small distances ($r<r_0$) the potential is the exact
Schwarzschild potential, and at great distances ($r>r_0$) the potential is
approximately Schwarzschild for $A=1$, and non-Schwarzschild for 
$A\ne 1$. (Toy potentials have been used before to study tail phenomena (in Schwarzschild \cite{price} and 
in RN and ERN \cite{bicak}), although not the same toy potential as in here.)

\begin{figure}[htbp]
  % \centering
   \includegraphics[width=3.4in]{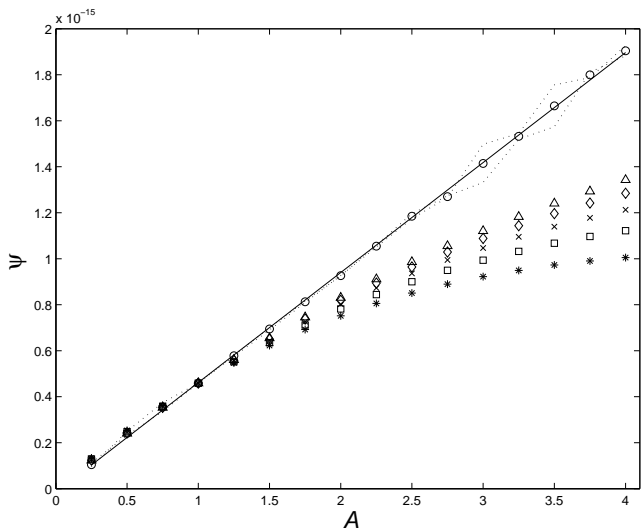} 
\caption{The magnitude of the tail in Schwarzschild at a fixed value of time as a
function of $A$ for different values of $r_0$. Same initial data  and
evaluation point were used in all cases. $\ast$: $r_0=100M$; 
$\square$: $r_0=125M$; $\times$: $r_0=150M$; $\diamond$: $r_0=175M$; $\triangle$: $r_0=200M$. The circles ($\circ$) are the Richardson extrapolations of the data at finite values of $r_0$ to $r_0\to\infty$, and the solid line is  a best fit line of the extrapolated data (circles). The squared correlation coefficient for the solid line is $R^2=0.9995$. The dotted curves are $3\sigma$ confidence curves of the extrapolated data.}
\label{amp_s}
\end{figure}

We next use a numerical code to study the dependence of the tails on
$A$ and $r_0$. The code we use, and convergence tests, are described 
in the Appendix. It is a Cauchy code with each cell computed by a characteristic ``diamond." The 
code is globally second-order convergent. We vary $r_0$ (in practice from $100M$ to $200M$), and for each value of $r_0$, we vary $A$ (in practice from $0.25$
to $4$). For each simulation (characterized by specific values of $r_0$ and
$A$) we verified that indeed tails are generated (in all cases with the
same tail exponent; see also Ref.~\cite{buzz}), and then we recorded the
value of the field in the tail regime for a fixed value of time. In
practice, we recorded the fields at $t=3000M$ at $r_*=0$, but the results do not depend
on the choice of the evaluation time or evaluation point. 
Figure \ref{amp_s} shows the
magnitude of the tail at a fixed time as a function of $A$ for different
values of $r_0$. It is clear from Fig.~\ref{amp_s} that the dependence of
the tail on $A$ is nonlinear for any {\it finite} value of $r_0$. However,
by increasing $r_0$ we find that the deviation from a linear dependence
decreases. To find the magnitude of the field in the limit as $r_0\to\infty$, we apply Richardson's deferred approach to the limit (``Richardson's extrapolation") for each value of $A$, and then fit the extrapolated data points to a straight line. Fig.~\ref{amp_s} suggests that the tail field would depend linearly on $A$ in the limit as $r_0\to\infty$. In fact, in
Fig.~\ref{amp_s} we look at successive simulations, in which we study the
dependence of the field on $A$ for scatterings occuring at $r>r_0$.

\begin{figure}[htbp]
  % \centering
   \includegraphics[width=3.4in]{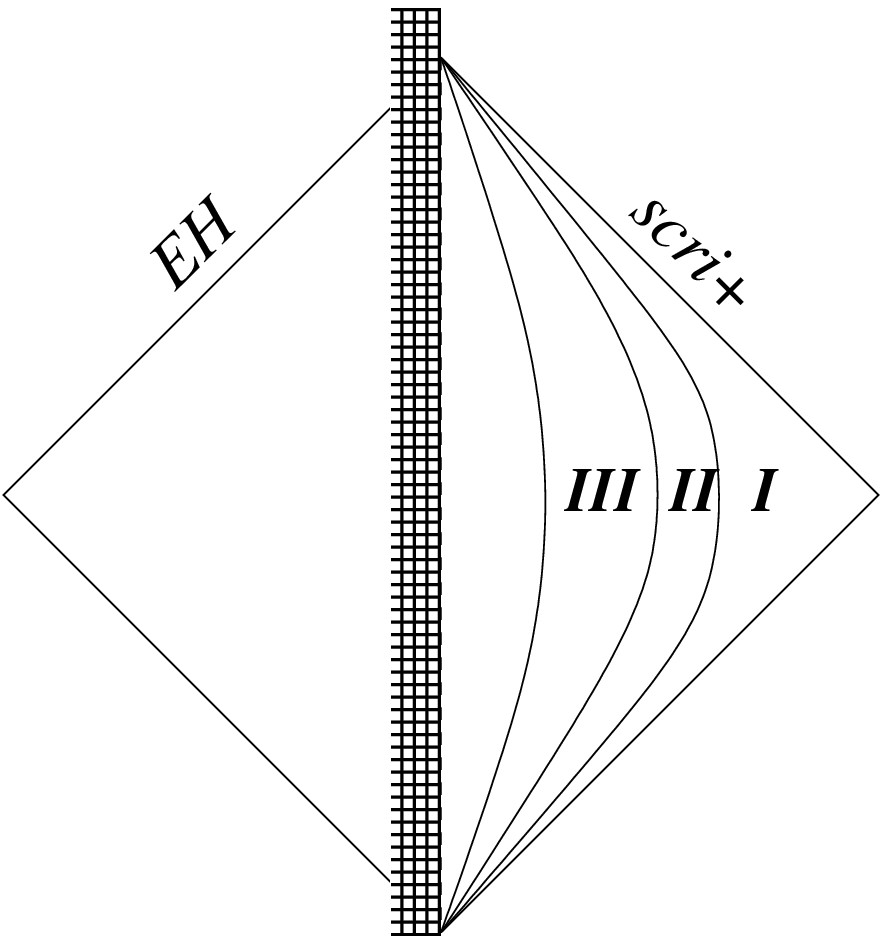} 
\caption{The spacetime diagram with three curves corresponding to three
different values of $r_0$, dividing spacetime into three regions I,II, and
III. See the text for more information.}
\label{regions}
\end{figure}

Figure \ref{regions} schematically shows the spacetime diagram with three
curves corresponding to three different values of $r_0$. For the smallest
value of $r_0$, the tails, as recorded at the observation point, have
contributions from scatterings at all values of $r>r_0$, i.e., from
regions III, II, and I in Fig.~\ref{regions}. For the intermediate value
of $r_0$ the tails have contributions from regions II and I, and for the
largest value of $r_0$ the contributions come only from region I
\cite{footnote}. 

Figure \ref{amp_s} suggests that when only scatterings at asymptotically  
large values of $r_0$ are considered, the dependence of the tail field on
$A$ would be linear, which implies that only secondary waves would be 
present. The tails are known to be the outcome of scatterings 
off the effective potential at asymptotically great distances, such that  
our results imply that indeed the tails in Schwarzschild at {\it
asymptotically} late times are caused by waves that are scattered just
once, i.e., secondary waves.

\begin{figure}[htbp]
  % \centering
   \includegraphics[width=3.4in]{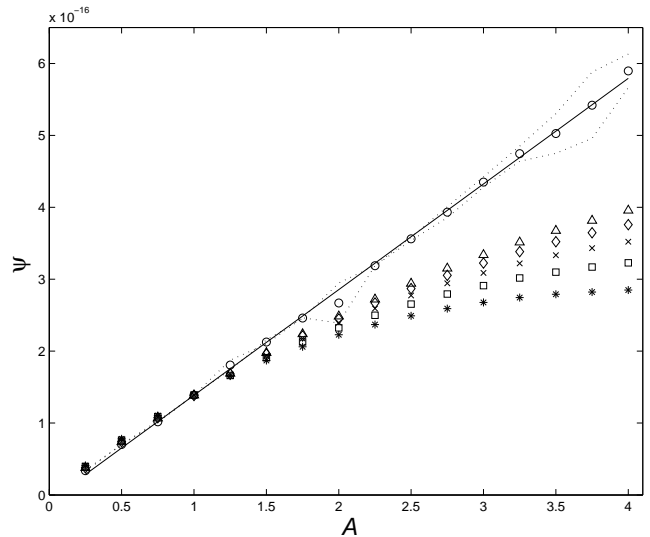} 
\caption{Same as Fig.~\ref{amp_s} for the ERN spacetime. The squared correlation coefficient for the solid line is $R^2=0.9987$. }
\label{amp_ern}
\end{figure}

Figure \ref{amp_ern} shows for the ERN spacetime the same information
as Fig.~\ref{amp_s} does for Schwarzschild. We infer that in ERN too, the
late-time tails are caused by secondary waves. In the ERN case the toy potential
was taken as
\begin{eqnarray}\label{toy2}
V(r_*)=\left\{\begin{array}[c]{l}
\left(1-\frac{M}{r}\right)^2\left[\frac{\ell(\ell +1)}{r^2}+
\frac{2M}{r^3}\left(1-\frac{M}{r}\right)\right] \; r<r_0 \\
\frac{\ell(\ell +1)}{r_*^2}+
A\frac{4M\ell(\ell +1)}{r_*^3}\, \ln [r_*/(2M)]  \;\;\;r>r_0\, .
\end{array}\right.
\end{eqnarray}

For both Fig.~\ref{amp_s} and Fig.~\ref{amp_ern} we used momentarily stationary initial data with compact support, but similar results were obtained also for other choices. In practice, we used initial data which are non-zero only between $r_{*{\rm i}}$ and $r_{*{\rm f}}$, where the field has the form
$$
\psi=\frac{[(r_*-r_{*{\rm i}})(r_*-r_{*{\rm f}})]^8}{[(r_{*{\rm c}}-r_{*{\rm i}})(r_{*{\rm c}}-r_{*{\rm f}})]^8}
$$
where $r_{*{\rm c}}=(r_{*{\rm i}}+r_{*{\rm f}})/2$, and $r_{*{\rm i}}=10M$, $r_{*{\rm f}}=30M$.

In fact, as for the Schwarzschild curvature potential $A=1$, it is the linearity of the field with $A$ in the neighborhood of unity which is important. Indeed, in both Figs.~\ref{amp_s} and \ref{amp_ern} the local linearity in the neighborhood of the physical curvature potential is implied.

\begin{figure}[htbp]
  % \centering
   \includegraphics[width=3.4in]{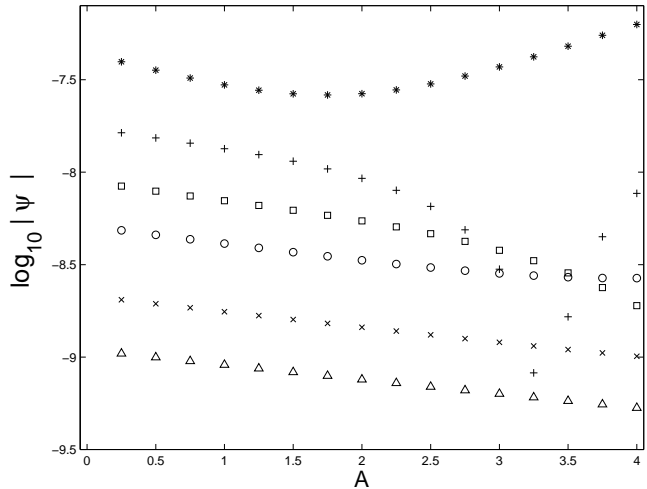} 
\caption{The magnitude of the tail in ERN at a fixed evaluation point as a
function of $A$ for different values of $t$, for the same choice of $r_0$. Same initial data  and
evaluation point ($r_*=0$) were used in all cases. Here, $\ell=1$, and $r_0=100M$. 
The evaluation times are:  
$\ast$: $t=300M$; $+$: $t=400M$;  $\square$: $t=500M$; $\circ$: $t=600M$; $\times$: $t=800M$; and 
$\triangle$: $t=1000M$. 
}
\label{AERNISM}
\end{figure}

The domination of secondary waves for the ERN tails for initial data that include a static moment is 
illustrated in Fig.~\ref{AERNISM}, that shows the field as a function of $A$ for a number of evaluation times at $r_*=0$. Here, the toy potential is 
as in (\ref{toy2}), except that the first equation in (\ref{toy2}) is taken for $r_*(r_0)>r_*>-r_{*}(r_0)$ and the second for $|r_*|>r_{*}(r_0)$.  
As shown in Fig.~\ref{AERNISM}, at early evaluation times the field is non-linear in $A$, and at later times it becomes linear in $A$. Also, for $A\sim 1$, the field is effectively linear  already at earlier times. Also for the case of an initially static moment present in the initial data, the conclusion is that the tails in ERN are governed by secondary waves. Again, also for early times (in the tail regime) the local linearity with $A$ in the neighborhood of the physical curvature potential is implied by Fig.~\ref{AERNISM}.

An important result is that for all $A\ne 0$, the tail power-law index remains 
unchanged. It is only the amplitude of the tail that is sensitive to $A$, not the decay rate. Same conclusions are found also for the case of initial data that include an initially static moment.

\section{ERN with initial data of compact support}\label{com}

In this Section we adapt the GPP argument to the case of ERN, when the
initial data have compact support. In Section \ref{ism} we consider the
case of initial data which include an initially static moment. 

Motivated by the preceding considerations, we define spacetime duality as follows:

\noindent {\bf Definition}: Two spacetimes (fixed background geometry and an intial data set for linearized perturbations) are said to be {\em dual} if the asymptotic contributions of secondary waves to the late-time tails in the two spacetimes (at a fixed observation point) have the same decay rate. 

\begin{figure}[htbp]
  % \centering
   \includegraphics[width=3.4in]{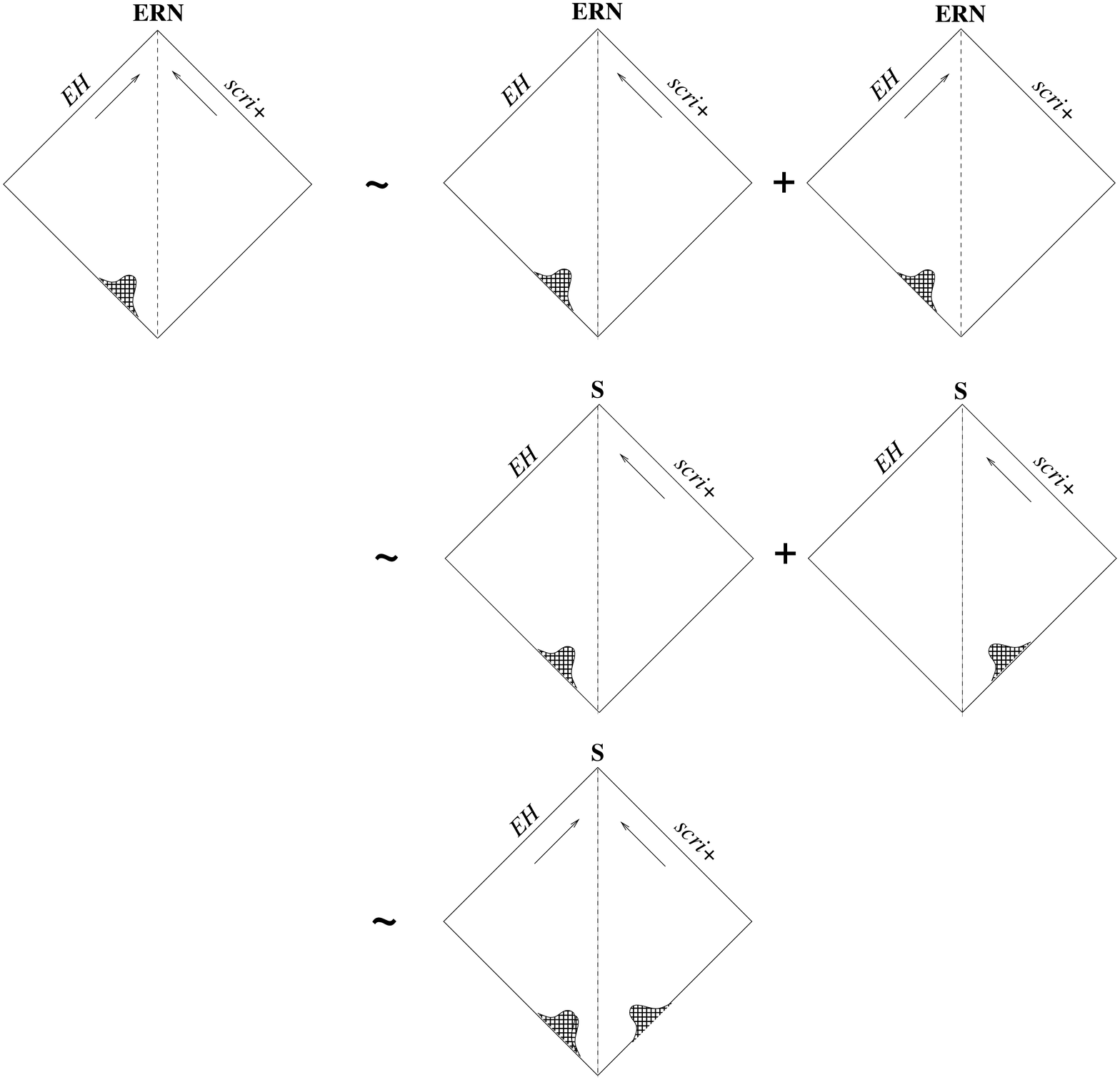} 
\caption{The duality transformations of the tail problem. See the
text for details.
}
\label{ern1}   
\end{figure}

Consider an ERN spacetime, with initial data of compact support. The tails
observed on $r={\rm const}$ at time $t$ have contributions ``coming from
the right'' (incoming waves) and contributions ``coming from the left''
(outgoing waves) (see the diagram on the left of
the top row of Fig.~\ref{ern1}). (We first consider the observation point
to coincide with the maximum of the effective potential for reasons to be
discussed below, but then our results for the tail are independent of the
evaluation point such that this assumption does not jeopardize the
generality of our discussion.) Because these are linear waves,
we may separate the waves on $r={\rm const}$ at time $t$ into the two
types of waves, incoming and outgoing, which make them. We next
argue, that the tails of the original problem (i.e., ERN with
perturbations of compact support, and waves which are a combination of
outgoing and incoming waves) are the same as the tails of two superposed
problems: ERN with the same initial data but only waves coming from the
right, and ERN with the same initial data but only waves coming from the   
left (right hand side of the top row in Fig.~\ref{ern1}). Such a
decomposition can be done because of two reasons: (i) the
linearity of the problem, and (ii) our previous result that it is only the
secondary waves which are important for the generation of the tails. (It
is important that we discuss only secondary waves. If higher-order waves
were allowed, then there would be no clear separation of waves coming from
the right and waves coming from the left, as the two would be coupled in
an intricate manner.)

Next, consider the case of ERN with the waves coming from the right. We
argue that the (partial) tails in this case are the same as the tails in
Schwarzschild for the same initial data, because at great distances the
effective potential in Schwarzschild is the same as in ERN. [Deviations
are only at $O(r_*^{-3})$.] Therefore, we can replace the first diagram on
the right hand side of the top row of Fig.~\ref{ern1} with the first
diagram on the second row. Similarly, the (partial) tails coming from the
left in ERN (second diagram on right hand side of the top row) are the
same as the tails in Schwarzschild, if we reflect the initial data to the
other side of the potential barrier, and change the initial data to be
initially incoming (second diagram on the second row). This is the case
because the effective potential in ERN close to the event horizon is the
same as the effective potential in ERN at great distances, and the latter
in its turn is the same as the effective potential at great distances in
Schwarzschild. Consequently, the ERN effective potential close to the
event horizon is asymptotically the same as the effective potential at
great distances in Schwarzschild. It is important that we reflect the
initial data because we need to keep the property of the original diagram,
that the direction of the waves is the same as the original direction of
the initial pulse. (The wave propagation is strictly speaking not the same
in these three spacetimes, because the effective potentials at finite
distances are not identical. However, the tails are generated only by the
asymptotic parts of the curvature potential, and these are the same in
these three spacetimes.) 

The original tail problem in ERN is then dual to the suporposition of two
Schwarzschild problems, which, because of the linearity of the problem,
we can recombine into a single problem by adding together the original
initial data and the reflected initial data. There is one more subtle
point, though: The tails we obtained are not the full tails, because in
the recombined diagrams we only have waves coming from the right, whereas
in the full picture in Schwarzschild we would have waves coming both from
the right and from the left. Adding the waves in Schwarschild which come
from the left, we argue, would not change the tails. Allowing for waves
coming from the left in Schwarzschild would add only negligible amount of
secondary waves, because in Schwarzschild the effective potential drops
off exponentially near the event horizon. In addition, because the 
original pulse has compact support, there will be no primary waves coming
from the left at late times. Therefore, we argue that we can add to our
diagram also the waves coming form the left in Schwarzschild, because they
will not change the tails. This way, we obtain the last diagram in
Fig.~\ref{ern1} (third row), which is the full diagram for a tail problem
in Schwarzschild for initial data different from the original one (the
original initial data superposed with their reflection). As the tail problem 
in Schwarzschild is completely understood (the exponent of the tails is
$2\ell+3$), we argue that the same exponent for the tails is expected also
in ERN. Below, in Section \ref{num}, we show, using numerical simulations,
that this is indeed the case.

\section{ERN with initial data with initial static moment}\label{ism}

The arguments in this case follow closely those of the preceding Section, and will therefore be described here briefly. For initial data with an ``initially static moment" (i.e., a multipole moment of the static solution is 
present on the initial data hypersurface), for Schwarschild 
$$\psi^{\rm Sch}=Q_{\ell}\left(\frac{r-M}{M}\right)\sim\frac{1}{r_*^{\ell +1}}\;\;\;\;\;\; (r\gg M)$$
where $Q_{\ell}$ is the Legendre function of the second kind, while for ERN \cite{bicak}
$$\psi^{\rm ERN}_+=\frac{1}{(r-M)^{\ell +1}}\sim \frac{1}{r_*^{\ell +1}}\;\;\;\;\;\; (r\gg M)$$
and
$$\psi^{\rm ERN}_-=(r-M)^{\ell}\sim \frac{1}{|r_*|^{\ell}}\;\;\;\;\;\; (0<r-M\ll M)\, .$$
Recall that the form of the initial data with an initially static moment is important only for $|r_*|\gg M$. That is, it is only the {\em asymptotic} drop off rate of the field (on the initial hypersurface) away from the peak of the effective potential that determines whether an initially static moment is present or not. Remarkably, the initially static moment initial data are the same for ERN as they are for Schwarzschild for $r_*\gg M$. We may therefore take the initial data in ERN to be $\psi^{\rm ERN}_+$ for $r_*>0$ and 
$\psi^{\rm ERN}_-$ for $r_*<0$. Notice that $\psi^{\rm ERN}_-$ is consistent with the dynamical requirement that a local observer sees a regular field as she is crossing the event horizon. This requirement is presented in \cite{bicak} as $\psi\sim {\rm const}+{\rm const}/u$ for all $\ell$, which at $t=0$ becomes 
$\psi\sim {\rm const}+{\rm const}/|r_*|$. This requirement yields, in fact, the {\em slowest} drop off of the field. Faster drop off is not disallowed, and will lead to a vanishing field on the EH, instead of a non-vanishing constant. Notice, however, that the dynamical requirement is consistent with our initial data (at $t=0$) only for $\ell=0,1$, such that for $\ell\ge 2$ it does not represent our choice for initial data. 

\begin{figure}[htbp]
  % \centering
   \includegraphics[width=3.4in]{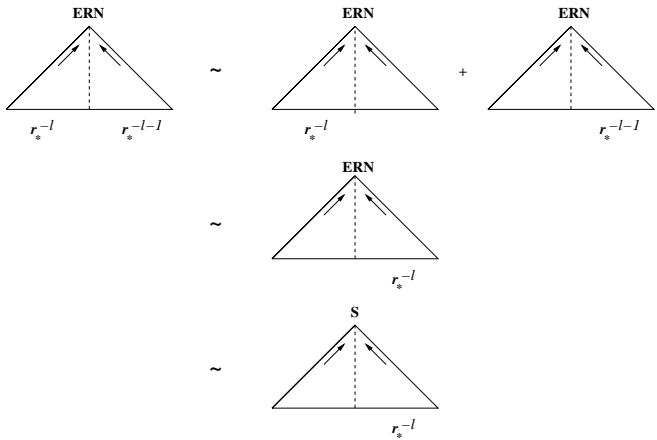} 
\caption{The duality transformations of the tail problem. See the
text for details.
}
\label{ern2}
\end{figure}

In Figure \ref{ern2} we describe the duality transformations as follows. First, the initial value problem, being a linear one, is equivalent to the superposition of two problems: in the first, 
$\psi=\psi^{\rm ERN}_+$ for $r_*>0$ and $\psi=0$ for $r_*<0$, and in the second, 
$\psi=0$ for $r_*>0$ and $\psi=\psi^{\rm ERN}_-$ for $r_*<0$. At the next step, because of the asymptotic symmetry of the ERN effective potential, the tail problem in ERN in the second problem above ($\psi=0$ for $r_*>0$ and $\psi=\psi^{\rm ERN}_-$ for $r_*<0$) is dual to an ERN problem with  
$\psi=\psi^{\rm ERN}_-$ for $r_*>0$ and $\psi=0$ for $r_*<0$. Because of the linearity of the scattering problem, we may now combine the two problems into a single one, for which 
$\psi=\psi^{\rm ERN}_-+\psi^{\rm ERN}_+$ for $r_*>0$ and $\psi=0$ for $r_*<0$. As for $r_*\gg M$ 
$\psi^{\rm ERN}_-\gg\psi^{\rm ERN}_+$, the initial data are dominated at large distances by $\psi^{\rm ERN}_-$, so that the tail problem is dual to ERN with $\psi=\psi^{\rm ERN}_-$ for $r_*>0$ and $\psi=0$ for $r_*<0$. Lastly, because of the duality of the tail problem in ERN and in Schwarzschild, the original tail problem is dual to a Schwarzschild tail problem, where the initial data are given by 
$\psi\sim r_*^{-\ell}$ for $r_*>0$ and $\psi=0$ for $r_*<0$.

\begin{figure}[htbp]
  % \centering
   \includegraphics[width=3.4in]{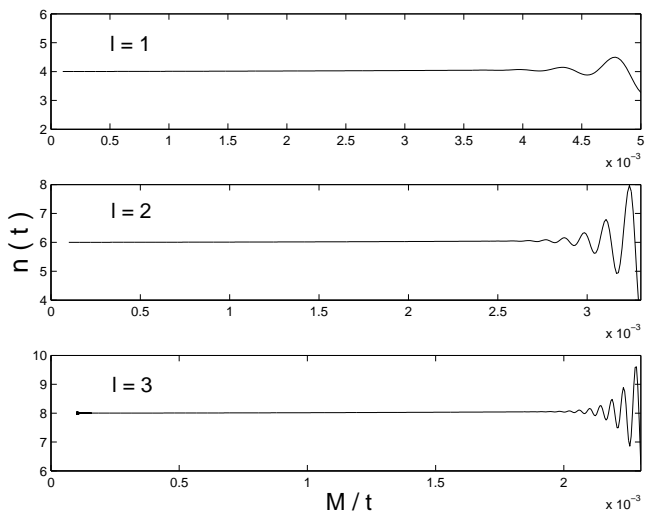} 
\caption{The local power index $n(t)$ as a function of $m/t$ in Schwarzschild, for initial data given by 
$\psi\sim r_*^{-\ell}$ for $r_*>0$ and $\psi=0$ for $r_*<0$, for a scalar field ($s=0$), for $\ell=1$ (upper panel), $\ell=2$ (middle panel), and $\ell=3$ (lower panel). See the
text for details.
}
\label{sch_ldata}
\end{figure}

The last tail problem is not a trivial one. It has initial data that are stronger than that of an initially static moment, but it is similar to the initially static moment in having non-compact initial data. A numerical solution of this problem, that is Schwarzschild with $\psi\sim r_*^{-\ell}$ for $r_*>0$ and $\psi=0$ for $r_*<0$ is presented in Fig.~\ref{sch_ldata}, which displays the local power index $n(t):=-t{\dot \psi}/\psi$ \cite{burko-ori} as a function of time for three values of $\ell$. 

%It is similar to a related problem with an initial static moment: Define $\ell'=\ell-1$. Then, the last problem has, for $r_*\gg M$, initial data of the form $\psi\sim r_*^{-\ell'-1}$. The curvature potential is 
%\begin{eqnarray}
%\hat V &=& 4M\,\frac{\ell(\ell+1)}{r_*^3}\,\ln\frac{r_*}{2M}\nonumber \\
%&=&4M\,A\,\frac{\ell'(\ell'+1)}{r_*^3}\,\ln\frac{r_*}{2M}\, .
%\end{eqnarray}
%where $A=(\ell'+2)/\ell'$
%This is a tail problem for an initial static moment with $\ell'$

\section{Numerical simulations of tails in ERN}\label{sec_num}

\begin{figure}[htbp]
  % \centering
   \includegraphics[width=3.4in]{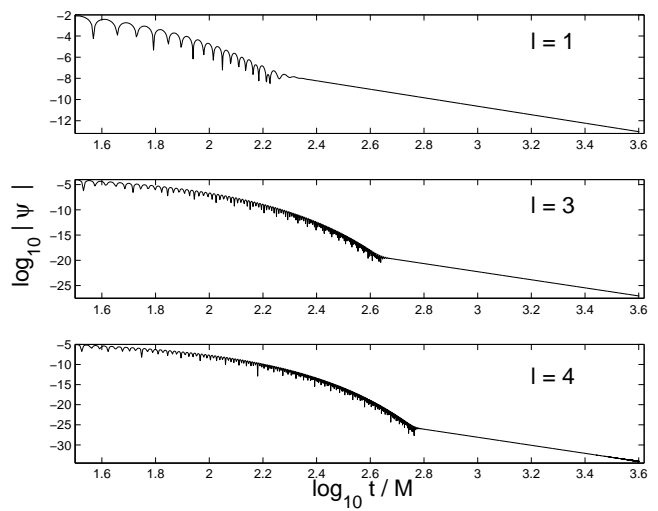} 
\caption{The field along $r_*=0$ as a function of $t$, for initial data with an initially static moment, for a scalar field ($s=0$), for $\ell=1$ (upper panel), $\ell=3$ (middle panel), and $\ell=4$ (lower panel). See the
text for details.
}
\label{ern_ism}
\end{figure}

\begin{figure}[htbp]
  % \centering
   \includegraphics[width=3.4in]{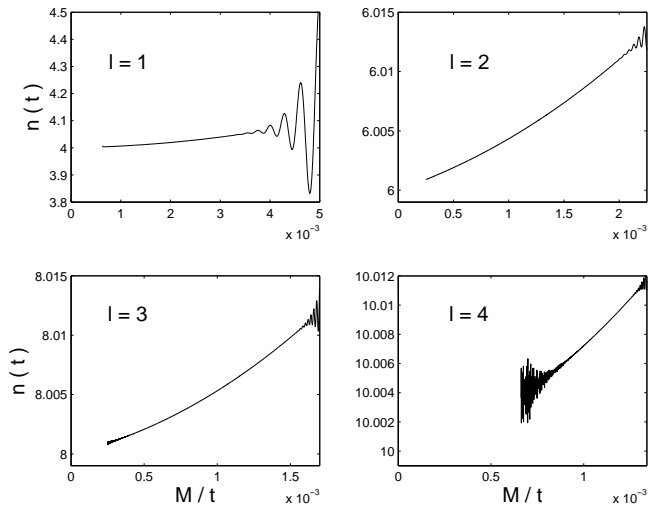} 
\caption{The local power index $n$ along $r_*=0$ as a function of $M/t$, for initial data with an initially static moment, for a scalar field ($s=0$), for $\ell=1$ (left upper panel), $\ell=2$ (right upper panel), $\ell=3$ (left lower panel), and $\ell=4$ (right lower panel). See the
text for details.
}
\label{ern_nt}
\end{figure}

The preceding discussion provides us with the expectation that initial data in ERN with an initial static moment present lead to the same indices of power-law tails as in Schwarzschild, namely at late times 
$\psi\sim t^{-(2\ell+2)}$. To test whether these expectations are indeed realized, we present in Fig.~\ref{ern_ism} the field along $r={\rm const}$ as a function of time for initial data of an initial static moment $\ell$, for various values of the latter. In Fig.~\ref{ern_nt} we present the local power index $n(t)$ for the same data. The numerical results agree with our prediction: the late time tails in ERN for initial data that are those of an initial static moment drop off with time as $\psi\sim t^{-(2\ell+2)}$. Finally, in Table \ref{comp} we confront our prediction for the tail power-law index in ERN with numerical results. The numerical results for the power-law indices are extrapolations to $t\to\infty$ of the local power indices $n(t)$. 

\begin{table}
 \caption{Confrontation of the prediction of this paper based on duality arguments ($2\ell+2$) for the power-law index along $r_*={\rm const}$ with our numerical results for ERN and initial data of an initially static moment. The relative error is  computed as the difference between the numerical result and $2\ell+2$, divided by the latter.}
  \centering
     \begin{tabular}{||c|c|c|c|c||} \hline
   $\ell$ {\bf  mode} & {\bf Prediction} & {\bf Numerical} & {\bf Relative}\cr
    & {\bf based on} & {\bf result} & {\bf error} \cr
     & {\bf duality} ($2\ell+2$) &  & 
   \cr \hline \hline
    $0$ & $2$ & $1.9972$ & $1\times 10^{-3}$ \cr \hline
    $1$ & $4$ & $3.9996$  & $1\times 10^{-4}$\cr \hline
    $2$ & $6$ & $5.9999$ & $2\times 10^{-5}$ \cr \hline 
    $3$ & $8$ & $7.9998$  & $3\times 10^{-5}$\cr \hline 
    $4$ & $10$ & $9.9974$  & $3\times 10^{-4}$\cr \hline  \hline
   \end{tabular}
\label{comp}
\end{table}

\begin{figure}[htbp]
  % \centering
   \includegraphics[width=3.4in]{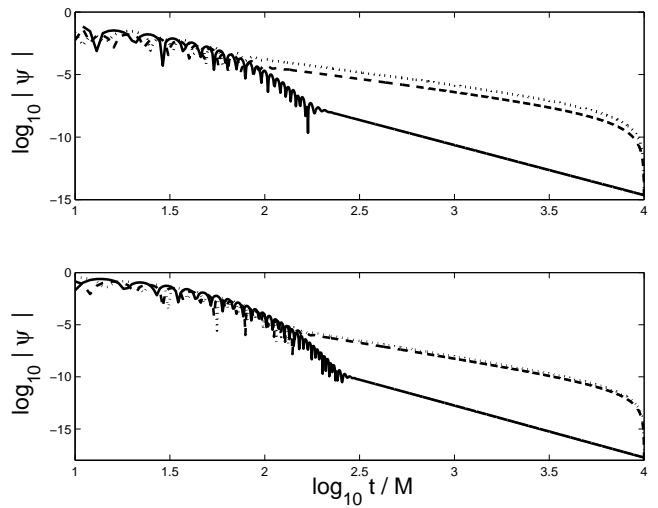} 
\caption{The field along $r_*=0$ (solid curve), along the event horizon (dotted curve), and along future null infinity (dashed curve). Upper panel: Initial data of a pure initially static moment. Lower panel: Initial data of an initially incoming pulse of compact support between $-10M<r_*<10M$. For both cases $\ell=1$. The ``bending" of the curves along the event horizon and along future null infinity for late advanced and retarded times, correspondingly, occurs because these two asymptotic regions of spacetimes are represented by large (but finite) retarded and advanced times, respectively. 
}
\label{EH_SCRI}
\end{figure}

The effective equivalence of the event horizon and future null infinity is further demonstrated by the tails along outgoing and incoming null rays at large values of advanced and retarded times, respectively. While the tails along a $r_*={\rm const}$ curve fall off at late times as $t^{-(2\ell+\mu+1)}$ for both Schwarzschild and ERN, where $\mu=1,2$ depending whether an initially static moment is present or not, along future null infinity the Schwarzschild tails fall off as $u^{-(\ell+\mu)}$, where $u=t-r_*$ is retarded time. We find this result to remain unchanged also for ERN (Fig.~\ref{EH_SCRI}). The main difference between Schwarzschild tails and ERN tails occurs along the event horizon: while in Schwarzschild the tails fall off as 
$v^{-(2\ell+\mu+1)}$, in ERN they fall off as $v^{-(\ell+\mu)}$, the same as along future null infinity (Fig.~\ref{EH_SCRI}). Here, $v=t+r_*$ is advanced time. 

\section*{Acknowledgments}
The authors thank Ji\v{r}\'{i} Bi\v{c}\'{a}k and Richard Price for discussions. CJB was supported by  a Research Experiences for Undergraduates in Science and Engineering fellowship, sponsored by the 
Alabama Space Grant Consortium, under Contract NNG05GE80H and by the UAH President's Office. LMB was supported in part by NASA/GSFC/EPSCoR grant No.~NCC--580.

\appendix

\section{Numerical code and Convergence tests}\label{app}

We have used two versions of the code, that allow us to use either Cauchy or characteristic data.   For the latter, we used a standard characteristic code in 1+1D in double-null coordinates $u,v$.  To use Cauchy data, we modified the code so that initial data (for $\psi$ and for $\dot\psi$) are specified at $t=0$, but each computational cell is calculated using the regular characteristic diamond. The computational domain is the domain of influence of the initial data (characteristic or Cauchy), so that no boundary conditions are specified. 

In what follows we denote $\psi_S=\psi(u,v)$, $\psi_E=\psi(u,v+\,\Delta v)$, $\psi_W=\psi(u+\,\Delta u,v)$, and $\psi_N=\psi(u+\,\Delta u,v+\,\Delta v)$. In each computational cell we use the second-order algorithm 
$$\psi_N=\psi_E+\psi_W-\psi_S-\frac{1}{4}\,V_0\,\psi_0\,\Delta u\,\Delta v\, .$$
In practice, we take $\psi_0=(\psi_E+\psi_W+\psi_N+\psi_S)/4$, so that we still need to solve for $\psi_N$. Collecting the coefficients of the fours field points, one finds that 
\begin{equation}\label{num}
\psi_N=\frac{1-\frac{1}{16}\,V_0\,\Delta u\,\Delta v}{1+\frac{1}{16}\,V_0\,\Delta u\,\Delta v}\,(\psi_E+\psi_W)-\psi_S\, .
\end{equation}
This scheme is second order. 

Characteristic data can now be evolved by straightforward marching. Cauchy data presents us with the problem of evolving the initial time step, as no field are specified on $\psi_S$. 
To determine the first time step according to the given initial data, consider first the case of momentarily stationary initial data. In that case, ${\dot\psi}=0$ at $t=0$, or $\psi_N=\psi_S$ on the initial slice.  Substituting into (\ref{num}), we find for momentarily stationary initial data 
\begin{equation}
\psi_N=\frac{1}{2}\,\frac{1-\frac{1}{16}\,V_0\,\Delta u\,\Delta v}{1+\frac{1}{16}\,V_0\,\Delta u\,\Delta v}\,(\psi_E+\psi_W) .
\end{equation}
Two other special cases to consider are $\psi_{,v}=0$ and $\psi_{,u}=0$. In the former case, $\psi_E=\psi_S$ on the initial slice, which we write as 
$\psi_S=-\psi_W+\psi_N+\psi_E$, so that  
\begin{equation}\label{in1}
\psi_N=\frac{\psi_W-\frac{1}{16}V_0\,\Delta u\,\Delta v\,\psi_E}{1+\frac{1}{16}\,V_0\,\Delta u\,\Delta v}\, .
\end{equation}
Similarly, if $\psi_{,u}=0$, one may take $\psi_W=\psi_S$ on the initial slice, or 
$\psi_S=\psi_W+\psi_N-\psi_E$, so that 
\begin{equation}\label{in2}
\psi_N=\frac{\psi_E-\frac{1}{16}V_0\,\Delta u\,\Delta v\,\psi_W}{1+\frac{1}{16}\,V_0\,\Delta u\,\Delta v}\, .
\end{equation}
General initial data for $\dot\psi$ can now be obtained by a linear combination of (\ref{in1}) and (\ref{in2}). In particular, the momentarily stationary initial data are obtained from (\ref{in1}) and (\ref{in2}) by averaging them. 

\begin{figure}[htbp]
  % \centering
   \includegraphics[width=3.4in]{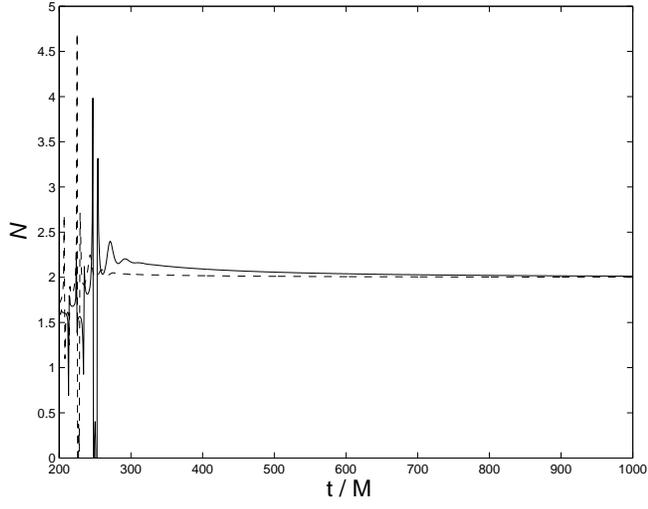} 
\caption{Convergence test for a scalar field ($s=0$) and dipole mode ($\ell=1$), 
for Schwarzschild (solid) and for ERN (dashed) for initial data with an initial static moment. Shown are the convergence order $N$ as a function 
of time along $r_*=0$. Oscillations of the convergence order at the quasi-normal mode epoch are typical of oscillatory data. 
}
\label{ct}
\end{figure}

We have done numerous convergence tests. First, we checked our code for a known exact solution (static solution in Schwarzschild). Then, we tested the global convergence order by finding the behavior of vector $L_p$ norms, and by monitoring the convergence order globally, throughout the entire  computational domain. As an illustration for the convergence tests we have done, Fig.~\ref{ct} shows the local convergence order along a $r_*={\rm const}$ worldline as a function of time. For grid spacings of $h,ah,a^2h$ we calculate the convergence order as
$$N=\log_{a}\left| \frac{\psi_{ah}-\psi_{h}}{\psi_{ah}-\psi_{a^2h}}\right| \, ,$$
where each of the three fields is evaluated at the same physical grid point. In all cases we found global second-order convergence throughout the computational domain. 

\begin{thebibliography}{99}
\bibitem{footnote} We note in passing that for any finite value of $r_0$
the field's amplitude appear to saturate at high values of $A$. This is an
interesting phenomenon which warrants further consideration.
\end{thebibliography}
\end{document}